\documentclass[aps, prb, superscriptaddress, twocolumn, letterpaper, 10pt, amsfonts, amsmath, amssymb]{revtex4-2}

\usepackage{graphicx}
\usepackage{hyperref} 
\usepackage{xcolor}
\usepackage{soul}
\hypersetup{
    colorlinks,
    linkcolor={blue!80!black},
    citecolor={blue!80!black},
    urlcolor={blue!80!black}
}
\usepackage{physics}

\usepackage{placeins}

\usepackage{siunitx}
\usepackage{chemformula}

\usepackage{xcolor}

\definecolor{AMColor}{rgb}{0.16, 0.23, 0.55}

\definecolor{editcolor}{rgb}{1, 0.0, 0.0}

\begin{document}
\title{Transient dynamics of a magnetic impurity coupled to superconducting electrodes: exact numerics versus perturbation theory}
\author{R. Seoane Souto}
\affiliation{Division of Solid State Physics and NanoLund, Lund University, S-22100 Lund, Sweden}
\affiliation{Departamento de F\'isica Te\'orica de la Materia Condensada, Condensed Matter Physics Center (IFIMAC) and Instituto Nicol\'as Cabrera, Universidad Aut\'onoma de Madrid, E-28049 Madrid, Spain}
\author{A. E. Feiguin}
\affiliation{Department of Physics, Northeastern University, Boston, Massachusetts 02115, USA}
\author{A. Mart\'in-Rodero}
\affiliation{Departamento de F\'isica Te\'orica de la Materia Condensada, Condensed Matter Physics Center (IFIMAC) and Instituto Nicol\'as Cabrera, Universidad Aut\'onoma de Madrid, E-28049 Madrid, Spain}
\author{A. Levy Yeyati}
\affiliation{Departamento de F\'isica Te\'orica de la Materia Condensada, Condensed Matter Physics Center (IFIMAC) and Instituto Nicol\'as Cabrera, Universidad Aut\'onoma de Madrid, E-28049 Madrid, Spain}

\date{\today}

\begin{abstract}
Impurities coupled to superconductors offer a controlled platform to understand the interplay between superconductivity, many-body interactions, and non-equilibrium physics. In the equilibrium situation, local interactions at the impurity induce a transition between the spin-singlet to the spin-doublet ground state, resulting in a supercurrent sign reversal ($0-\pi$ transition). In this work, we apply the exact time-dependent density matrix renormalization group method to simulate the transient dynamics of such superconducting systems. We also use a perturbative approximation to analyze their properties at longer times. These two methods agree for a wide range of parameters. In a phase-biased situation, the system gets trapped in a metastable state characterized by a lower supercurrent compared to the equilibrium case. We show that local Coulomb interactions do not provide an effective relaxation mechanism for the initially trapped quasiparticles. In contrast, other relaxation mechanisms, such as coupling to a third normal lead, make the impurity spin relax for parameter values corresponding to the equilibrium $0$ phase. For parameters corresponding to the equilibrium $\pi$ phase the impurity converges to a spin-polarized stationary state. Similar qualitative behavior is found for a voltage-biased junction, which provides an effective relaxation mechanism for the trapped quasiparticles in the junction.
\end{abstract}

\maketitle

\section{Introduction}
Superconductors are macroscopic coherent materials, described by a complex order parameter $\Delta$, measuring the pairing amplitude between electrons. They are characterized by a non-dissipative electric current through the material for sufficiently low bias voltages \cite{Bardeen_PR1957}. In conventional superconductors, the electron pairing induces a gap in the density of states at the Fermi level with a value $\abs{\Delta}$. However, a spatial modulations of $\Delta$ can give rise to discrete excitation energies inside the gap due to the multiple Andreev reflections \cite{DeGennes_PL1963,Andreev_JETP1964,Andreev_JETP1966,Kulik_JETP1970}. These states have been observed in superconductors coupled to other materials,
including semiconductors, normal metals, and other superconductors, for recent reviews see \cite{Sauls_review2018,Prada_NRP2020}. 

The coupling between superconductors and magnetic impurities or small quantum dots offers the possibility to investigate the competition between superconductivity and many-body interactions \cite{Cuevas_PRB2001,Balatsky_RMP2006,Jarillo_Nat2006,Eichler_PRL2007,Grove_NPJ2007,Eichler_PRB2009,Franceschi_Nat2010,Rodero_Adv2011,Meden_IOP2019}. Coulomb blockade tends to fix the number of electrons in the system, suppressing charge fluctuations and blocking the current through the system \cite{Beenakker_book}. In the strong Coulomb blockade limit, the ground state of an impurity with an odd number of electrons is doubly degenerate, hosting an unpaired spin \cite{Balatsky_RMP2006}. In this regime, spin fluctuations increase the low-bias conductance, known as the Kondo effect \cite{Goldhaber_Nat1998,Kouwenhoven_PW2001}, characterized by the Kondo temperature, $T_K$.

The competition between Kondo and pairing correlations lead to a  phase diagram where the system transits between a singlet to a doublet ground state \cite{Rozhkov_PRL1999,Rozhkov_PRB2000,Vecino_PRB2003,Tanaka_NJP2007}. At the transition between the doublet and the singlet states, the subgap states (also known as Yu-Shiba-Rusinov states) cross the Fermi level \cite{Pillet_Nat2010,Dirks_Nat2011,Jellinggaard_PRB2016}, reversing supercurrent direction, the so-called $0-\pi$ transition \cite{Glazman_JETP1989,Clerk_PRB2000,vanDam_Nat2006,Cleuziou_Nat2006,Jorgensen_NanoLett2007,Delagrange_PB2018}. At the ground state transition, BCS pair correlations are suppressed and the electron pairing is predominantly spin-triplet close to the impurity \cite{Kuzmanovski_PRB2020,Perrin_PRL2020}.

In equilibrium at zero bias voltage, this problem has been studied using numerically exact methods such as numerical renormalization group \cite{Yoshioka_JPSJ2000,Choi_PRB2004,Oguri_JPSJ2004,Karrasch_PRB2008,Rodero_JPCM2012,Domanski_SR2016,Grove_NatCom2018,Zalom_PRB2021,Moca_PRL2021}, quantum Monte Carlo \cite{Siano_PRL2004,Luitz_PRB2010,Luitz_PRL2012,Pokorny_PRR2020}, and also real-time diagrammatic methods \cite{Governale_PRB2008}. In the regime of weak Coulomb interaction (in comparison to the gap and the tunnel coupling to the leads), the problem is well-described by the lowest-order expansion in the Coulomb strength \cite{Shiba_PTP1973,Rozhkov_PRL1999,Rodero_JPCM2012,Wrzesniewski_arXiv2020}. The resulting mean-field approximation breaks the spin-rotation symmetry for sufficiently strong Coulomb interaction. Higher-order corrections in the expansion, describing quantum fluctuations, cures this shortcoming \cite{White_PRB92,Souto_NJP2018}. A second-order description succeeds in describing the $0-\pi$ transition and observables in the $0$ phase \cite{Vecino_PRB2003,Zonda2015,Zonda_PRB2016}.

The time dynamics of impurities coupled to superconductors can answer fundamental questions, like the time onset of electron correlations, formation of subgap states, dependence on the initial conditions, and relaxation of quasiparticles. These questions are inaccessible from stationary calculations. They are essential for controlling the internal state of the system, relevant for quantum information applications \cite{Mishra_arXiv2021}, and to study the interplay between superconductivity and many-body interactions under non-equilibrium conditions, including the possible excitation of the Higgs mode \cite{Shimano_review2020}.

In the absence of electron-electron interactions, the transient dynamics of a superconducting nanojunction has been studied in Refs. \cite{Perfetto_PRB2009,Souto_PRL16,Souto_PRB17,Taranko_PRB2018,Taranko_PRB2019,Wrzesniewski_arXiv2020,Averin_PRB2020,Taranko_PRB2021}. In this case, the system is not able to thermalize due to the trapping of quasiparticles in the subgap states \cite{Stefanucci_PRB2010,Souto_PRL16,Souto_PRB17}. These quasiparticles are long-lived, being an obstacle towards the efficient control and manipulation of superconducting devices \cite{Zgirski_PRL11,Janvier_science15}.

The effect of electron correlations has been much less studied in the non-equilibrium case \cite{Avishai_PRB03,Eichler_PRL2007,Dell'Anna_PRB08,Andersen_PRL2011,Avriller_PRL15,Tuovinen_NJP2021} and the transient regime \cite{Kamp_PRB2021,Heckschen_arXiv2021}. In this work, we develop an exact numerical method that simulates the real-time evolution of the system, based on the time-dependent density matrix renormalization group (tDMRG). This method is a powerful numerical technique to solve the time-dependent Schr\"odinger equation in finite systems, accounting for many-body interactions \cite{White2004a,Feiguin2005,Heidrich_PRB2009,vietri,Feiguin2013b,Paeckel2019}. To study the long-time properties, we use a second-order perturbation expansion on the Coulomb strength, describing the system dynamics for a wide range of parameters.


We show that local interactions do not provide an effective relaxation mechanism for the trapped quasiparticles in the subgap states. In contrast, the inclusion of a third normal lead or a finite bias allows for quasiparticle relaxation due to transitions to excited states with energies above $\abs{\Delta}$ \cite{Yeyati_PRL2003}. In this last case, the superconducting phase difference evolves linearly in time, giving rise to a time oscillating current due to multiple Andreev reflections (MAR) \cite{Arnold_LTP1987,Averin_PRL1995,Bratus_PRL1995,Cuevas_PRB96,Scheer_PRL1997,Scheer_Nature1998,Ludoph_PRB2000,Goffman_PRL2000,Yeyati_PRB97,Souto_PRB17,Villas_PRB2020}. We show that the impurity converges to a state with spin-rotation symmetry for parameters in the $0$ phase. In contrast, it converges to a spin-polarized state for parameters in the $\pi$ phase and an initially trapped spin in the impurity.

The rest of the manuscript is organized as follows: In Sec. \ref{Sec::model} we introduce the model and the two methods used in our study: tDMRG and perturbative Green functions. In Sec. \ref{Sec::results} we present the main results of our work. We first focus on the equilibrium situation, Sec. \ref{SubSec::eq}, where we compare both methods and analyze the convergence to the steady-state. In Sec. \ref{SubSec::Thermalization} we analyze the question of thermalization in the case where a metallic lead couples weakly to the impurity. The voltage-biased situation is discussed in Sec. \ref{SubSec::V}, where we show that it provides an effective relaxation mechanism for the initially trapped quasiparticles. Additional results complementing the ones shown in the main text can be found in the Appendix. Finally, we present the main conclusions and a brief outlook in Sec. \ref{Sec::Conclusions}.


\section{Model and formalism}
\label{Sec::model}
\subsection{Model}
We consider an impurity coupled to superconducting electrodes, described by the Hamiltonian
\begin{equation}
    H=H_{i}+H_{leads}+H_{T}(t)\,.
\end{equation}
The impurity is described by a spin-degenerate level
\begin{equation}
    H_{i}=\varepsilon_0\hat{\Psi}^{\dagger}_0 \hat{\tau}_z \hat{\Psi}_0+Un_{\uparrow}n_{\downarrow}\,,
    \label{H_i}
\end{equation}
where $\Psi_0=(d_{0\uparrow},d^{\dagger}_{0\downarrow})^T$ is the Nambu spinor, with $d_{0\sigma}$ being the annihilation operator for an electron with spin $\sigma$ at the impurity, and $\hat{\tau}$ are the Pauli matrices in the Nambu space. Here, $\varepsilon_0$ is the impurity energy level, $U$ the charging energy, and $n_{\sigma}=d^{\dagger}_{0\sigma}d_{0\sigma}$, with $\sigma=\uparrow,\,\downarrow$. The leads are described as a 1-dimensional chain of electronic sites, given by
\begin{equation}
    H_{leads}=\sum_{\nu,j}\left(\hat{\Psi}^{\dagger}_{\nu j} \hat{h}_{\nu j} \hat{\Psi}_{\nu j}+t_0\hat{\Psi}^{\dagger}_{\nu j} \hat{\tau}_{z} \hat{\Psi}_{\nu j-1}\right)\,,
\end{equation}
where $t_0$ is the inter-site chain hopping parameter, $h_{\nu j}=\xi_\nu \hat{\tau}_z+\Delta_\nu\hat{\tau}_x$, with $\xi$ being the onsite energy of electron at site $j$, and $\Delta_\nu$ the superconducting gap with $\nu=L,R$.

The hopping between the impurity and the leads is described by
\begin{equation}
    H_{T}(t)=\hat{\Psi}^{\dagger}_{\nu 0} \hat{V}_{\nu}(t)\hat{\Psi}_{0}+\mbox{H.c.}\,.
    \label{H_T_def}
\end{equation}
Here, $\hat{V}_{\nu}(t)=\theta(t)V_\nu\hat{\tau}_z e^{i\hat{\tau}_z\phi_\nu(t)/2}$, where we have considered the tunneling amplitude, $|V_\nu|$, to be momentum-independent. The tunneling rate to the leads is given by $\Gamma_\nu=\pi|V_\nu|^2\rho_\nu$, where $\rho_\nu=1/\pi t_0$ is the normal density of states at the Fermi level, and $\Gamma=\Gamma_L+\Gamma_R$. We consider the impurity to be suddenly connected to the leads at $t=0$. The superconducting phase is $\phi_\nu(t)=\phi_\nu(0)+\mu_\nu t$, where $\mu_\nu$ is the chemical potential of $\nu$ electrode. In the following we consider a symmetric voltage drop in the junction, $\mu_L=-\mu_R=V/2$ with $V$ being the applied bias voltage.

In this work we focus on the time-evolution of single particle observables. The spin-polarization of the impurity is given by $ S_Z(t)=[ n_\downarrow(t)-n_\uparrow(t)]/2$, where
\begin{equation}
    \label{Eq_n_sigma}
    n_\sigma(t)=\left\langle d_{0\sigma}^\dagger(t) d_{0\sigma}(t)\right\rangle\,,
\end{equation}
is the occupation per spin and $n_0(t)=n_\uparrow(t)+n_\downarrow(t)$ the impurity charge. The current at $\nu$ interface is given by
\begin{equation}
    \label{Eq_I}
     I_\nu(t)=i\,\mbox{Tr}\left[\,\hat{V}_{\nu}(t)\left\langle \hat{\Psi}^{\dagger}_{\nu 0}(t)\hat{\Psi}_{0}(t)\right\rangle\right]+\mbox{H.c.},
\end{equation}
where the trace is taken in the Nambu space.
We define the symmetrized current as $\left\langle I\right\rangle=\left( I_L-I_R\right)/2$. Finally, the electron pair amplitude at the impurity is given by
\begin{equation}
    \label{Eq_Delta}
    \Delta_i(t)=\Delta\left\langle d_{0\uparrow}(t)d_{0\downarrow}(t)\right\rangle.
\end{equation}

\subsection{Time-dependent density matrix renormalization group}

The method tDMRG is a powerful numerical technique to solve the time-dependent Schr\"odinger equation in finite systems with a great degree of control and numerical precision. In the same way as in its ground state version, tDMRG relies on a matrix product state representation of the wave function plus a Suzuki-Trotter decomposition of the evolution operator to evolve the state \cite{White2004a,Feiguin2005,Heidrich_PRB2009,vietri,Feiguin2013b,Paeckel2019}. Among other applications, it has been used to study time-dependent transport in a number of one-dimensional setups \cite{Schmitteckert2004,Al-Hassanieh2006a,Al-Hassanieh2008a,Boulat2008,dasilva2010,heidrich-meisner2010,Rincon2014,Schlunzen_PRB2017,Petrovic2021,petrovic2021b,Braganca_PRB2021}. The formulation is extensively discussed in the cited literature, so we limit ourselves to a brief summary of the protocol used in this work. Typically, in all the situations discussed here, we are interested in the evolution of the the system after it is perturbed at time $t=0$ by some external potential (a quench of the tunneling amplitudes, for instance). We usually observe that the wave function develops a wave-front that propagates away from the center, translating into a transient regime. After some characteristic time that depends on the model parameters, observables such as the current reach a pseudo-steady-state, equilibrating around a constant value. Eventually, the wave front hits the boundaries of the open system, the wave packet bounces back, and the current is reversed. In order to extend the duration of the steady-state time window, we delay the wave packet at the edges by introducing ``damped boundary conditions'' with exponentially decaying hopping amplitudes near the boundaries of the system, slowing down the electrons near the edges. \cite{dasilva2008,Feiguin2008f}. We use a hoping that decays exponentially with the distance to the impurity as $t_j=t_0 \Lambda^{-j/2}$ between $j$ and $j+1$ sites in the left and right leads. For our calculations, we use a chain with $N=120$ sites and $\Lambda=20$, which we found optimal for convergence.

In this work, we extend the tDMRG method to the case where the leads are superconducting, including a local pairing term $\Delta d_{n\uparrow}d_{n\downarrow}+\mbox{H.c.}$ at every site of the chain, except at the impurity. A particular consideration to take into account about the included pairing term is that the particle number is conserved {\it modulo two} \cite{Feiguin2007a} ({\it i.e.}, number parity and spin are good quantum numbers). To preserve unitarity and prevent loss of accuracy, we pick a bond dimension (the dimension of the DMRG basis) large enough to keep the truncation error under a predetermined tolerance, and we stop the simulations either when the accuracy grows beyond this value, or when the effects of the boundaries become dominant. We typically use a bond dimension of $200$ and a tolerance of $10^{-6}$. The observables we represent are given by Eqs. (\ref{Eq_n_sigma}-\ref{Eq_Delta}) for the impurity population, current, and pair amplitude.

Our protocol is implemented as follows: before calculating the evolution, we prepare the system in its global ground state. In this work, we focus on the situation where the impurity is initially detached from the leads. The initial impurity occupation is controlled by the initial level energy, $\varepsilon_0(t=0^-)$. We also introduce an initial Zeeman splitting in the impurity Hamiltonian \eqref{H_i} of the form
\begin{equation}
    h_z(t=0^-)(n_{\uparrow}-n_{\downarrow})\,.
\end{equation}
At the initial time, $t=0$, the impurity is connected to the superconducting leads and the impurity Hamiltonian takes the form of Eq. \eqref{H_i}, with $h_z=0$ and $\varepsilon_0$ assuming its final value. We then evolve the wave function using tDMRG. Finally, we include the bias voltage as a time-dependent hoping between the impurity and the leads, using the expression in Eq. \eqref{H_T_def}.  

\subsection{Non-equilibrium Green functions}
\label{Sec::NEGFs}
Perturbation theory on the interaction strength, $U$, provides accurate results in the $U/\Gamma\lesssim1$ regime. This approximation has been shown to describe accurately physical observables in the stationary regime and the non-magnetic phase \cite{Vecino_PRB2003,Zonda2015,Zonda_PRB2016}. In this work, we extend this approximation to the time domain, allowing us to describe the non-stationary and non-equilibrium situations. For this task, we extend to the superconducting case, the self-consistent formalism introduced in Refs. \cite{Souto_NJP2018,Souto_book}, describing the formation of the Kondo resonance at the Fermi level for metallic electrodes. The method is based on a time discretization of the non-equilibrium Green functions (NEGFs) \cite{Tang_PRB2014,Tang_PRB2014b,Souto_PRB2015,Avriller_PRB2019}. The impurity Green function is given by
\begin{equation}
    \hat{G}_0=\left(\hat{g}^{-1}_0-\hat{\Sigma}_{leads}-\hat{\Sigma}_{int}\right)^{-1}\,.
    \label{GreenFunction}
\end{equation}
We solve the problem in a discrete time mesh, where the inverse of the isolated impurity Green function is given by 
\begin{equation}
    \hat{g}^{-1}_0=\begin{pmatrix}
g^{-1}_e & 0\\
0 & g^{-1}_h
\end{pmatrix}\,.
\end{equation}
The electron and hole parts of the inverse Green function in the discrete mesh are described in Refs. \cite{Souto_PRL16,Souto_PRB17}.

The leads self-energy is given by \cite{Rodero_Adv2011}
\begin{equation}
    \hat{\Sigma}^{\alpha\beta}_{leads}(t,t')=\alpha\beta\sum_\nu \hat{V}_\nu(t)\hat{g}^{\alpha\beta}_{\nu}\hat{V}^{\ast}_\nu(t')\,,
\end{equation}
where $\alpha$, $\beta=\pm$ denote the Keldysh branch. The leads boundary Green functions are given by
\begin{equation}
    \hat{g}^{\alpha\beta}_{\nu}(\omega)=\begin{pmatrix}
\mathfrak{g}^{\alpha\beta}_\nu(\omega) & \mathfrak{f}^{\alpha\beta}_\nu(\omega)\\
\mathfrak{f}^{\alpha\beta}_\nu(\omega) &  \mathfrak{g}^{\alpha\beta}_\nu(\omega)
\end{pmatrix}\,.
\end{equation}
The bare Green functions for a superconducting chain are given by

\begin{equation}
    \mathfrak{g}^{r/a}_\nu(\omega)=\frac{\omega\pm i0^+}{2t^{2}_0}\left[1-\frac{\sqrt{W^2-(\omega\pm i 0^+)^2}}{\sqrt{\Delta^2-(\omega\pm i0^+)^2}}\right]
\end{equation}
and
\begin{equation}
    \mathfrak{f}^{r/a}_\nu(\omega)=-\Delta/(\omega\pm i0^+)\mathfrak{g}^{r/a}_\nu(\omega)
\end{equation}
with $W=\sqrt{4t_{0}^2+\Delta^2}$ being the bandwidth and $0^+$ an infinitesimal. The Keldysh components are given by $\hat{g}^{+-}_\nu(\omega)=[\hat{g}^{a}_\nu(\omega)-\hat{g}^{r}_\nu(\omega)]n_F(\omega)$ and $\hat{g}^{-+}_\nu(\omega)=[\hat{g}^{a}_\nu(\omega)-\hat{g}^{r}_\nu(\omega)][n_F(\omega)-1]$, where $n_F$ is the Fermi distribution function. 

To calculate the interaction self-energy, we perform a perturbation in $U/\Gamma$. Up to second order in the perturbative parameter, the self-energy is given by $\hat{\Sigma}_{\rm int}=\hat{\Sigma}_{\rm \rm HF}+\hat{\Sigma}^{(2)}$. The linear term in $U$ is the mean-field approximation, given by
\begin{equation}
    \begin{split}
    \left[\Sigma_{HF}(t,t')\right]^{\alpha\beta}_{nn}=U\alpha\, n_{\bar{n}\bar{n}}(t)\,\delta(t,t')\delta_{\alpha\beta}\,,\\
    \left[\Sigma_{HF}(t,t')\right]^{\alpha\beta}_{n\bar{n}}=U\alpha\, n_{n\bar{n}}(t)\,\delta(t,t')\delta_{\alpha\beta}\,,
    \end{split}
\end{equation}
where $n=1,2$ denote the Nambu component and $\bar{n}$ the opposite component to $n$. Here, $n_{nn}$ denotes the charge per spin, given by $n_\uparrow=n_{11}(t)=-iG^{+-}_{11}(t,t)$ and $n_\downarrow=n_{22}(t)=-iG^{-+}_{22}(t,t)$. The other two components, off-diagonal in Nambu space, describe the induced pairing at the impurity at the mean field level, $n_{n\bar{n}}(t)=-iG^{+-}_{n\bar{n}}(t,t)$. The pairing amplitude in the impurity, $\Delta_i$ is given by the $n_{12}(t)$ component.

Electron correlation effects are described by higher order terms in the expansion. They contain information about the electron-hole pair creation due to the local Coulomb repulsion. Up to second order in $U/\Gamma$, the self-energy is given by
\begin{widetext}

\begin{equation}
    \begin{split}
    \left[\Sigma^{(2)}(t,t')\right]^{\alpha\beta}_{nn}&=-U^2\sum_{j=1,2}(-1)^{j-n}\left[G_{HF}(t,t')\right]_{nj}^{\alpha\beta}\left[G_{HF}(t,t')\right]_{\bar{n}j}^{\alpha\beta}\left[G_{HF}(t',t)\right]_{\bar{n}\bar{n}}^{\beta\alpha}\,,\\
    \left[\Sigma^{(2)}(t,t')\right]^{\alpha\beta}_{n\bar{n}}&=U^2\sum_{j=1,2}(-1)^{j-n}\left[G_{HF}(t,t')\right]_{nj}^{\alpha\beta}\left[G_{HF}(t,t')\right]_{\bar{n}j}^{\alpha\beta}\left[G_{HF}(t',t)\right]_{n\bar{n}}^{\beta\alpha}\,,
    \end{split}
\end{equation}
\end{widetext}
where we have perturbed over the mean-field Green function, given by
\begin{equation}
    \hat{G}_{HF}=\left(\hat{g}^{-1}_0-\hat{\Sigma}_{leads}-\hat{\Sigma}_{HF}\right)^{-1}\,.
    \label{G_HF}
\end{equation}
This approximation has been shown to provide accurate results in the normal case, describing the onset of the Kondo peak at temperatures lower than $T_K$ \cite{White_PRB92,Karrasch_JCMP2008,Uimonen_PRB2011,Souto_NJP2018}.

The system observables can be obtained from the Green's function \eqref{GreenFunction}. The average charge and spin at the impurity are given by $n_0=n_\uparrow+n_\downarrow$ and $S_Z=(n_\uparrow-n_\downarrow$)/2. The current at a given interface can be calculated as \cite{Vecino_PRB2003}
\begin{equation}
    I_\nu=\mbox{Tr}\left\{\tau_z\left[\hat{V}_\nu(t) \hat{G}^{+-}_{\nu0}(t,t)-\hat{V}^{\ast}_\nu(t) \hat{G}^{+-}_{0\nu}(t,t)\right]\right\},\
\end{equation}
where the trace is taken over the Nambu space.

\section{Results}
\label{Sec::results}
\subsection{Phase-biased junction}
\label{SubSec::eq}
\begin{figure}
    \centering
    \includegraphics[width=1\columnwidth]{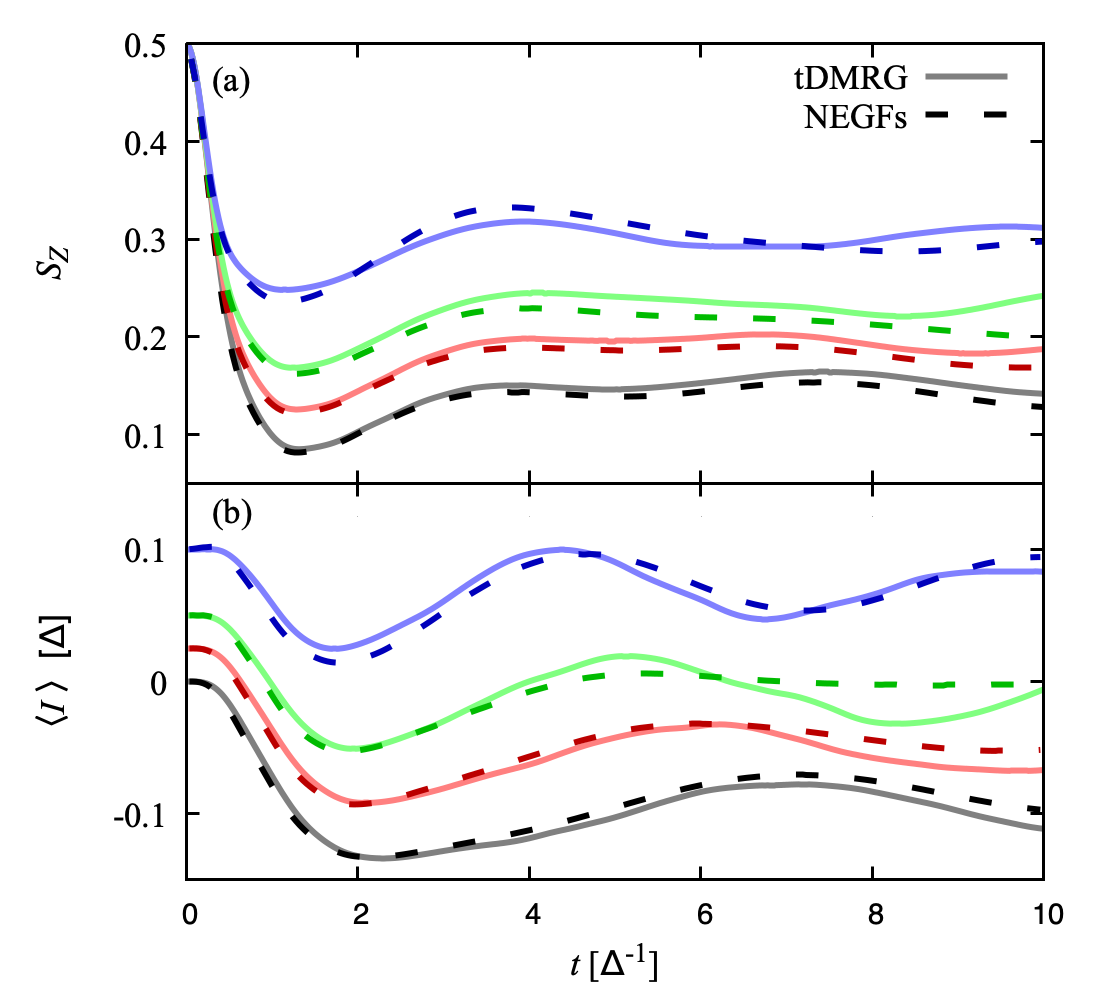}
    \caption{Comparison between tDMRG (solid line) and perturbative NEGFs (dashed line) methods. We show the time evolution of the impurity spin, $S_Z$ (a) and the symmetrized current (b) for $U/\Gamma=$0 (black), 1 (red), 2 (green), and 4 (blue). Current curves are shifted upwards by $U/40$ for clarity. The remaining parameters are $\Gamma=\Delta= 0.2 t_0$, $\phi=0.6\pi$, $\varepsilon_0=-U/2$ and the initial impurity occupation $(n_\uparrow,n_\downarrow)=(0,1)$.}
    \label{Fig:comparison_n01}
\end{figure}
In the phase-biased situation ($V=0$), a supercurrent can flow through the junction if the two leads have different superconducting phases. In the stationary regime, the current at zero temperature is given by the phase derivative of the ground state energy as a function of the phase difference \cite{Annett_book}. In Fig. \ref{Fig:comparison_n01} we show results for the current and the spin after a sudden connection of the impurity to the leads. We compare results for the two methods developed in our work, tDMRG and perturbative NEGFs, for an initial spin in the impurity, {\it i.e.} $(n_\uparrow(0),n_\downarrow(0))=(0,1)$. In the absence of interactions ($U=0$), both methods provide similar results. The small differences found, mostly visible at long times, are due to the renormalization of the electron pairing amplitude in the leads close to the impurity. This effect is taken into account within tDMRG but absent in the perturbative NEGFs approach. Both methods agree for a relatively wide range of $U/\Gamma$ values. In Fig. \ref{Fig:comparison_n01} we show results for $S_Z$ and the current through the system. In App. \ref{Appendix:pair} we also display the time evolution of the pairing potential at the impurity. The results of both methods start to deviate for $U/\Gamma\approx6$, predicting different stationary values for $S_Z$ (not shown).

In the limit $\Delta\gg T_K$, the two methods  describe the trapping of a magnetic moment in the impurity at long times \cite{Souto_PRL16,Souto_PRB17}. We have checked that no spin relaxation happens for times $t\lesssim50/\Delta$ (the steady-state value has been reached at $t\sim10/\Delta$). This is an indication that localized Coulomb interactions do not provide an effective relaxation mechanism for the initially trapped spin in the impurity. On the contrary, localized Coulomb interactions tend to quench the short-time spin relaxation, leading to a higher $S_Z$ stationary value when increasing $U$, Fig. \ref{Fig:comparison_n01} (a). The absence of thermalization causes a supercurrent reduction with respect to the expected value at equilibrium \cite{Souto_PRL16}.



\begin{figure}
    \centering
    \includegraphics[width=1\columnwidth]{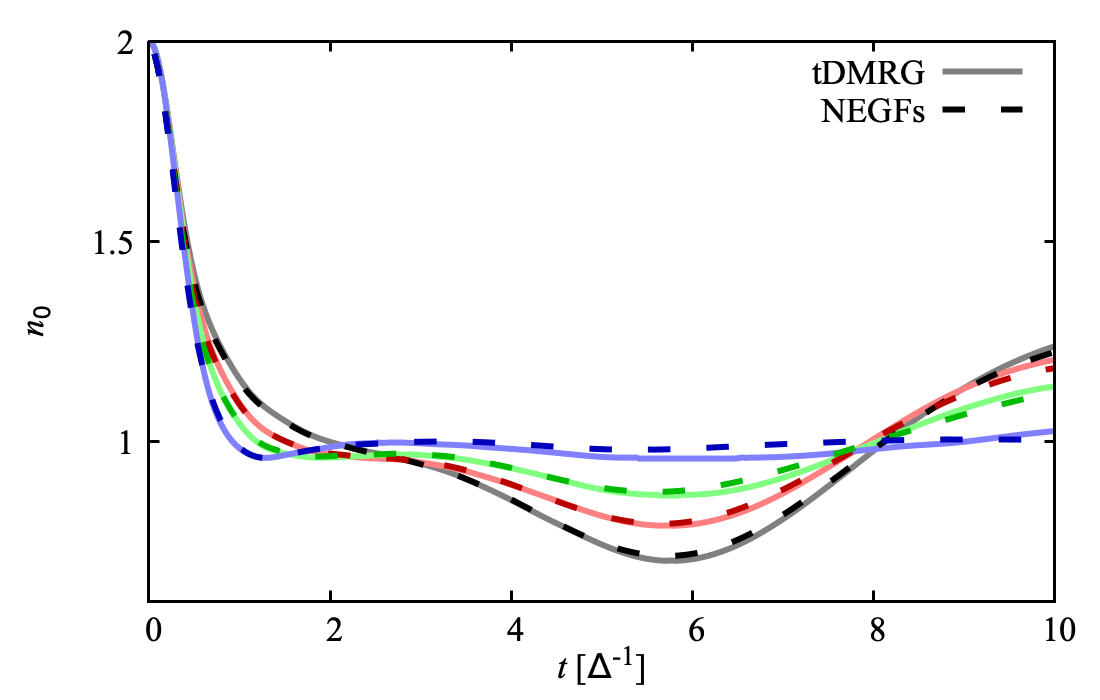}
    \caption{Evolution of the impurity charge for an initially fully occupied level, $(n_\uparrow(0),n_\downarrow(0))=(1,1)$ obtained using tDMRG (solid lines) and perturbative NEGFs (dashed lines). The remaining parameters are the same as in Fig. \ref{Fig:comparison_n01}: $\Gamma=\Delta=0.2t_0$, $\phi=0.6\pi$, $\varepsilon_0=-U/2$, and $U/\Gamma=0$ (black), 1 (red), 2 (green), and 4 (blue).}
    \label{Fig:population_n00}
\end{figure}

In the case of a fully occupied or empty impurity level, the charge exhibits undamped oscillations \cite{Souto_PRB17,Taranko_PRB2018}, shown in Fig. \ref{Fig:population_n00}. The oscillation frequency is given by the energy of the subgap states formed at the interface between the impurity and the superconductors (analytic expressions for the oscillations are given in Ref. \cite{Souto_PRB17} for $U=0$). The amplitude of the oscillations depend on the interaction strength, which tends to decrease its height. This is due to the increased energy difference between the ground state, with $n_0=1$, and the excited states with two and zero electrons. In the limit $U/\Delta\gg1$, the oscillations quenched and $n_0$ converges to the expected stationary state.

\subsection{Thermalization}
\label{SubSec::Thermalization}

\begin{figure}
    \centering
    \includegraphics[width=1\columnwidth]{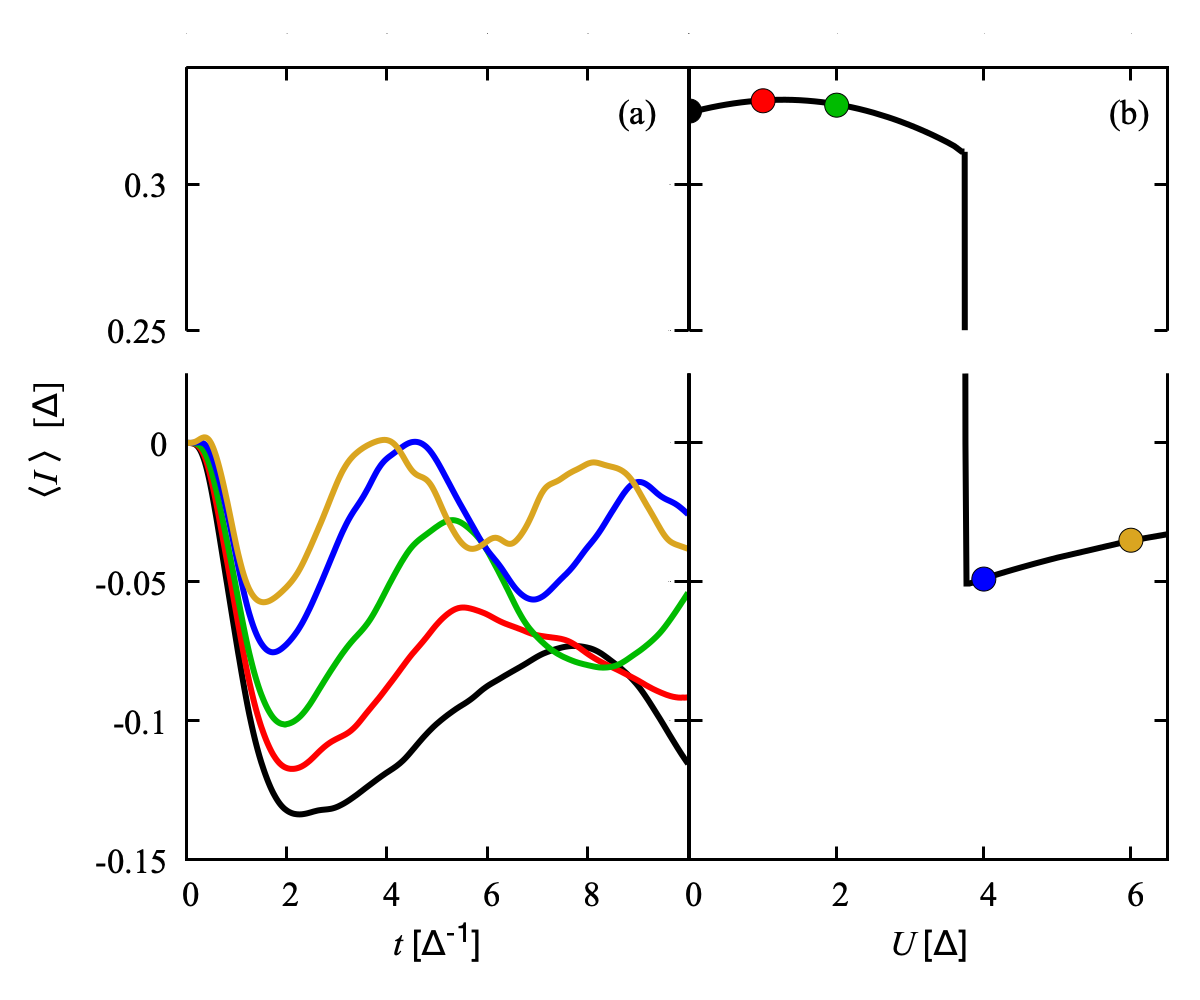}
    \caption{(a) Time-dependent current calculated using tDMRG for the same parameters shown in Fig. \ref{Fig:comparison_n01}: $\Gamma=\Delta=0.2t_0$, $\phi=0.6\pi$, and $\varepsilon_0=-U/2$. (b) Stationary current calculated using tDMRG, where the colored dots indicate the value of $U$ in the left panel.}
    \label{Fig:t_vs_state}
\end{figure}

The results in Fig. \ref{Fig:comparison_n01} and \ref{Fig:population_n00} suggest the trapping of long-lived quasiparticles in the system. Their effects are better illustrated in the long-time current, shown in Fig. \ref{Fig:t_vs_state} for the exact tDMRG method. The supercurrent converges to a lower steady-state, Fig. \ref{Fig:t_vs_state} (a), compared to the equilibrium result, Fig. \ref{Fig:t_vs_state} (b). The difference is better observed before the supercurrent sign reversal at $U_c/\Gamma\sim3.7$. The lower supercurrent value is due to the trapping of quasiparticles in the subgap states, which have an infinite lifetime within this model. Therefore, the long-time state is dominated by the short-time dynamics, when the Andreev bound states are formed \cite{Souto_PRL16}.

The absence of thermalization is consistent with the observed long relaxation times for trapped quasiparticles observed in experiments \cite{Zgirski_PRL11,Janvier_science15}. The finite relaxation time in these systems could be due to photon or phonon emission/absorption processes \cite{Olivares_PRB2014} which are not included in the present analysis. In addition, a small normal density of states inside the gap can provide a finite quasiparticle lifetime. Finally, transitions to the continuum states forced by an applied bias can allow quasiparticles to diffuse into the bulk of the superconductor, making the junction relax. The obtained results suggest that local Coulomb interactions do not provide by themselves an effective relaxation mechanism for trapped quasiparticles in the junction. This might not be the case if one considers extended interactions in the leads \cite{Ljubotina_arXiv2021}.

\begin{figure}
    \centering
    \includegraphics[width=1\columnwidth]{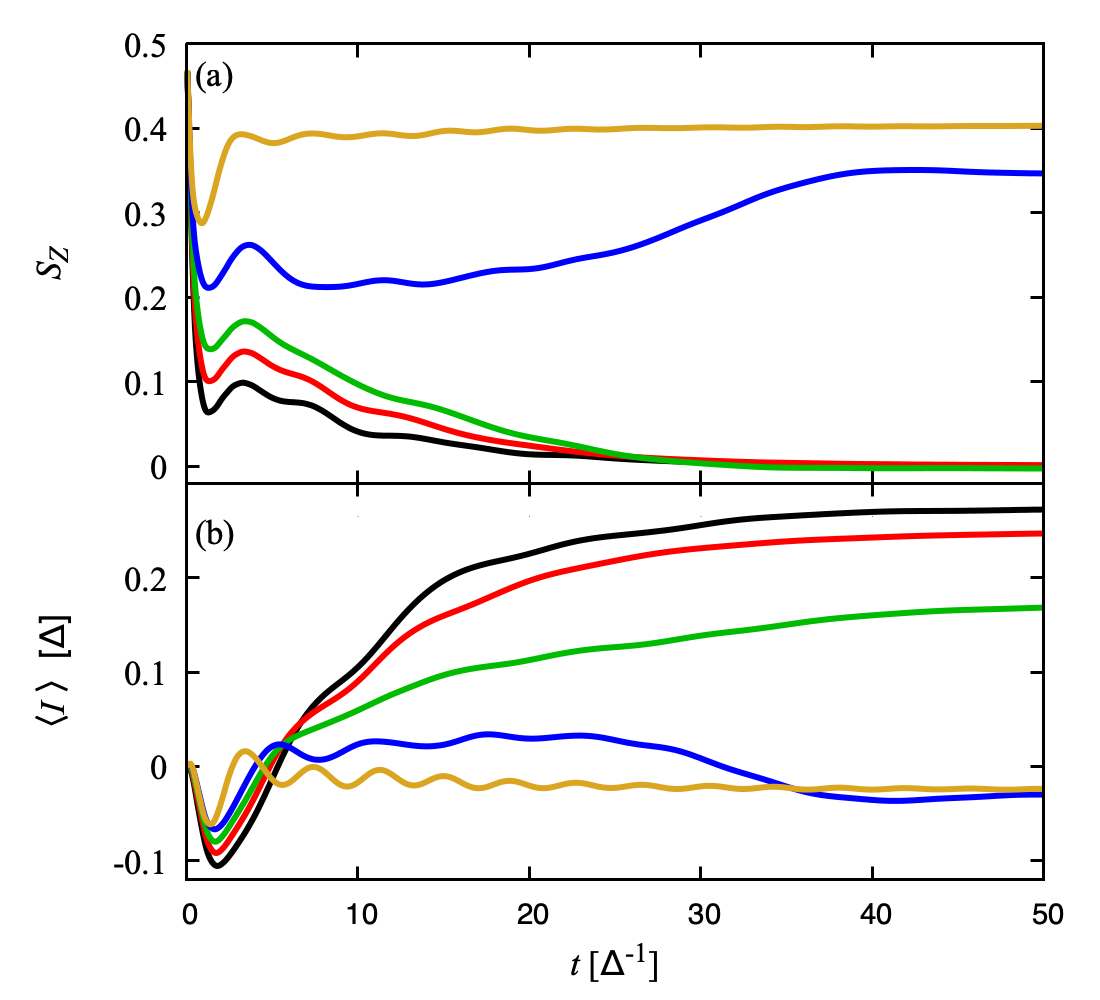}
    \caption{(a) Evolution of the impurity spin polarization using NEGFs. We tunnel couple the impurity to a normal lead, with a tunneling amplitude  $\eta/\Delta=0.1$. For parameters in the $0$ phase, $U/\Gamma=$0 (black), 1 (red), and 2 (green), the impurity spin converges to an unpolarized stationary state. In contrast, $S_Z$ converges to a finite value for parameters in the $\pi$ phase, $U/\Gamma=$4 (blue) and 6 (yellow). (b) Current evolution for the same cases shown in panel (a). The remaining parameters are the same as in Fig. \ref{Fig:comparison_n01}: $\Gamma=\Delta=0.2t_0$, $\phi=0.6\pi$, and $\varepsilon_0=-U/2$.} 
    \label{Fig:fig4}
\end{figure}

To simulate the mechanisms commented above for quasiparticle relaxation, we attach an additional normal lead to the impurity \cite{Souto_PRL16,Taranko_PRB2018,Taranko_PRB2019}, Fig. \ref{Fig:fig4}. In this case, the supercurrent exhibits a sign change when increasing $U/\Gamma$, characteristic of the $0-\pi$ transition, Fig. \ref{Fig:t_vs_state} (b). We find a smaller stationary value with respect to the exact calculations in Fig. \ref{Fig:t_vs_state} (b), due to the finite tunneling coupling to the normal electrode, $\eta$. The relaxation rate is given by $\eta$ when the system is far away from the $0-\pi$ transition. In contrast, the system relaxation close to the transition ($U_c/\Gamma\approx 3.7$ for this choice of parameters) becomes slower, with a relaxation time exceeding $1/\eta$. This behavior is more clearly seen in panel (a), where we represent $S_Z$. It indicates that perturbations take longer to relax close to the transition. This phenomenology is similar to the critical slowing down close to a phase transition. However, in our case, the $0-\pi$ transition becomes smooth due to the finite $\eta$ chosen. Therefore, it is not possible to observe the expected divergence of the relaxation timescale at the transition.


The additional relaxation mechanism considered makes the spin in the impurity converge to different stationary values, depending on the spin of the ground state, Fig. \ref{Fig:fig4} (a). For the smallest $U/\Gamma$ values shown, $S_Z$ converges to zero, consistent with expected singlet ground state. The situation is different in the $\pi$ phase, where $S_Z$ converges to a non-zero value for an initial condition with broken spin-rotation symmetry. Similar stationary results have been obtained for different $\eta$ values. In the next subsection, we show that a finite bias voltage provides a similar qualitative behavior, which acts as an effective relaxation mechanism for quasiparticles trapped in the subgap states.

\subsection{Voltage-biased junction}
\label{SubSec::V}
A bias voltage applied to the junction produces a linear increase of superconducting phase difference in time as $\phi(t)=\phi(0)+V\,t$. For $V < 2\Delta$, multiple Andreev reflections lead to the transference of multiple charges between the superconductors, resulting in a time-oscillating as well as a dc steady-state current. In the low-bias situation ($V\ll \Delta, \Gamma$), the subgap states can adapt instantaneously to the phase change and an adiabatic picture is valid. The situation is more complex in the case where $V$ is comparable to $\Delta$ and $\Gamma$. In this situation, non-adiabatic effects are important and higher harmonics have to be considered \cite{Cuevas_PRB96}.

\begin{figure}
    \centering
    \includegraphics[width=1\columnwidth]{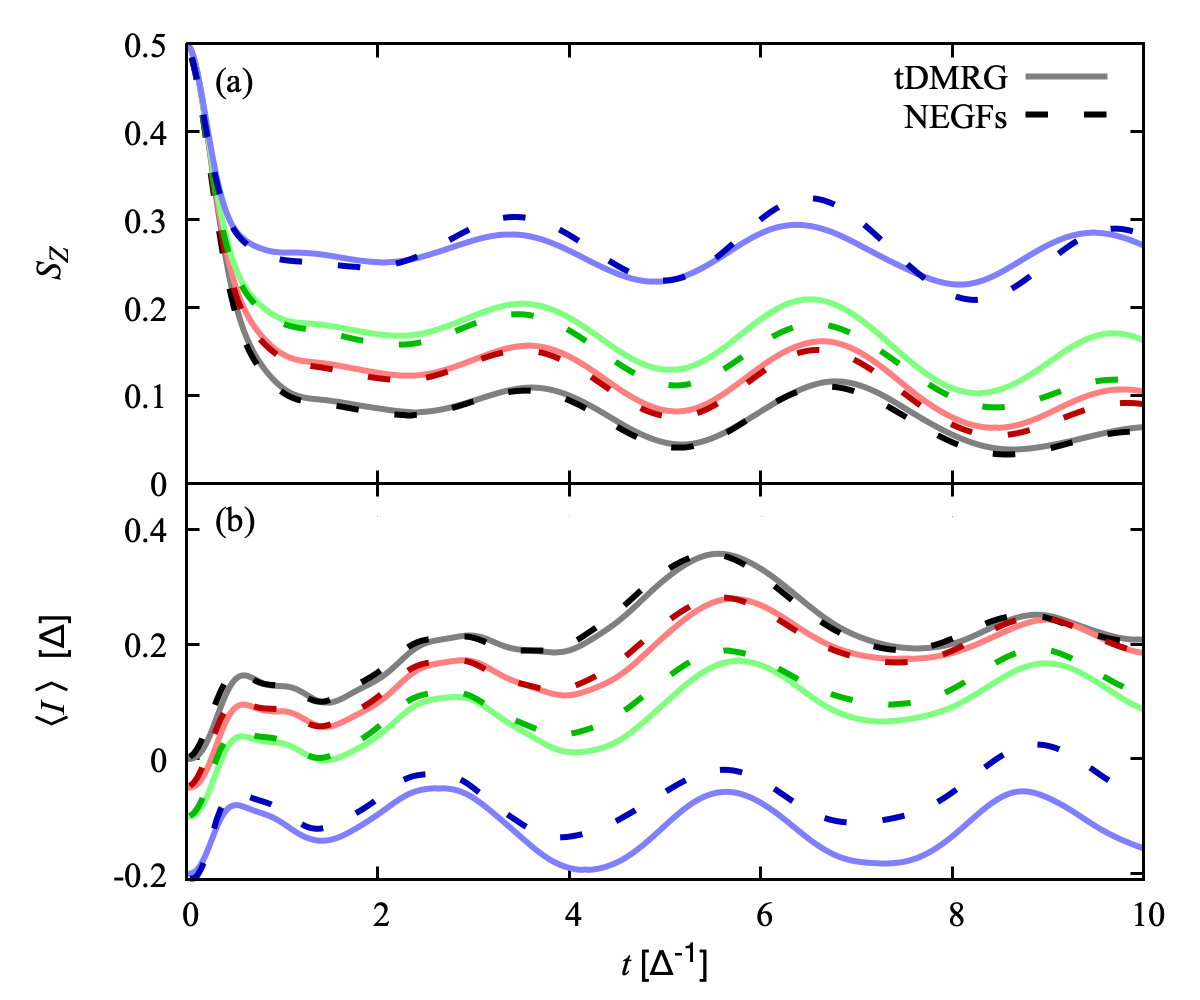}
    \caption{Spin polarization (a) and current (b) for a voltage-biased junction with $V=\Delta$ and $U/\Gamma=$0 (black), 1 (red), and 2 (green), and 4 (blue). The remaining parameters are the same as in Fig. \ref{Fig:comparison_n01}: $\Gamma=\Delta=0.2t_0$, $\phi=0.6\pi$, and $\varepsilon_0=-U/2$. Curves in panel (b) have been shifted down by $U/20$ for clarity.}
    \label{Fig:comparison_n01_V}
\end{figure}

In Fig. \ref{Fig:comparison_n01_V} we compare the results from tDMRG and the perturbative NEGFs for $V=\Delta$. Similarly to the phase-biased situation studied before, both methods agree for $U/\Gamma\lesssim1$. They start to deviate for the largest $U/\Gamma$ values shown, where the differences are more evident in the current evolution. We show additional calculations for different $V$ values in App. \ref{Appendix:comparison_V}. We additionally compare the pairing amplitude,  $\Delta_i$ in App. \ref{Appendix:pair}, exhibiting time-dependent oscillations with the same period as $S_Z$.

\begin{figure}
    \centering
    \includegraphics[width=1\columnwidth]{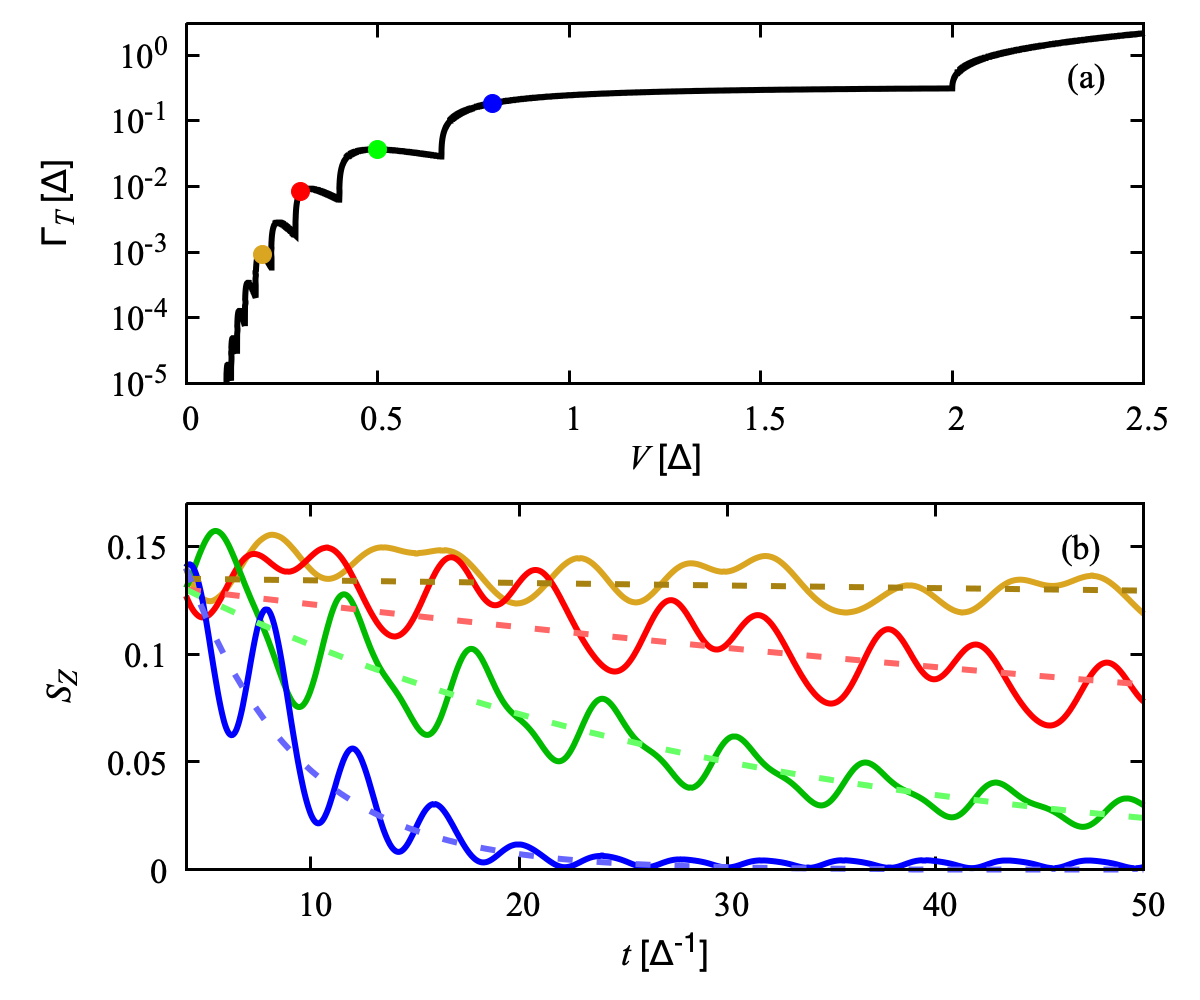}
    \caption{(a) Analytic relaxation rate of a voltage-biased junction given in Eq. \eqref{gamma_T}. (b) Relaxation of the initially trapped spin, comparing NEGFs numerical results (solid lines) and the analytic rates (dashed lines). We show results for $U/\Gamma=0$ and $V/\Delta=0.2$ (yellow), $0.3$ (red), $0.5$ (green) and $0.8$ (blue) from top to bottom, see colored dots in panel (a). The remaining parameters are the same as in Fig. \ref{Fig:comparison_n01}: $\Gamma=\Delta=0.2t_0$, $\phi=0.6\pi$, and $\varepsilon_0=0$.}
    \label{Fig:fig6}
\end{figure}

An applied bias voltage provides a relaxation mechanism for the trapped quasiparticles, making the initial condition relax, as shown in the upper panel of Fig. \ref{Fig:comparison_n01_V} for the smallest $U/\Gamma$ values. This behavior is better illustrated at longer times. In Fig. \ref{Fig:fig6} we show the time evolution of $S_Z$ for an initially trapped spin for the perturbative NEGFs method described in Sec. \ref{Sec::NEGFs}. In this case, the local magnetic moment of the impurity converges to the expected stationary value ($S_Z=0$ for a system with spin-rotation symmetry) at long times, Fig. \ref{Fig:comparison_n01_V} (a). 

In this case, relaxation is due to transitions between the subgap states and the continuum of states at energies above $\abs{\Delta}$. In the limit where the bandwidth $W\gg\Delta,\Gamma$, the transition rate is given by \cite{Yeyati_PRL2003}
\begin{equation}
    \Gamma_T=2\sum_{n\geq0}J_{n}^2(\tilde{\Delta}/V)\mbox{Im}\left[\Sigma(nV+V/2)\right]\,,
    \label{gamma_T}
\end{equation}
where $J_n$ are the Bessel functions of order $n$, $\tilde{\Delta}$ the induced gap at the impurity, and $\Sigma(\omega)=\omega\sqrt{\Delta^2-\omega^2}/\Gamma$. Eq. \eqref{gamma_T} predicts a relaxation rate that decreases with the energy gap between the subgap and the continuum of states, $\tilde{\Delta}$. It increases with the voltage, showing steps at $V=2\abs{\Delta}/(2n+1)$, Fig. \ref{Fig:fig6} (a). At these voltages, a new MAR channel involving the excitation of a single electron to the continuum  of states opens. The applied bias voltage leads to an exponential relaxation of the initially trapped spin in the impurity, \ref{Fig:fig6} (b). We represent the function $Ae^{-\gamma_T t}$ with dashed lines, describing the long-time relaxation rate of the impurity. Here, $\gamma_T$ is given by Eq. \eqref{gamma_T} and $A$ is taken as the only free parameter in the fit.

For $\Gamma/\Delta\ll1$, the subgap states appear close to zero energy exhibiting a small modulation with $\phi$. In this case, the relaxation rate is suppressed for $V<\Delta$. In the opposite regime, $\Gamma/\Delta\gg1$, known as the quantum point contact limit, the energy difference between the subgap and the continuum of states tends to zero. In this situation, the relaxation happens in a much faster timescale, determined the ac Josephson period for any voltage \cite{Souto_PRB17}. 

\begin{figure}
    \centering
    \includegraphics[width=1\columnwidth]{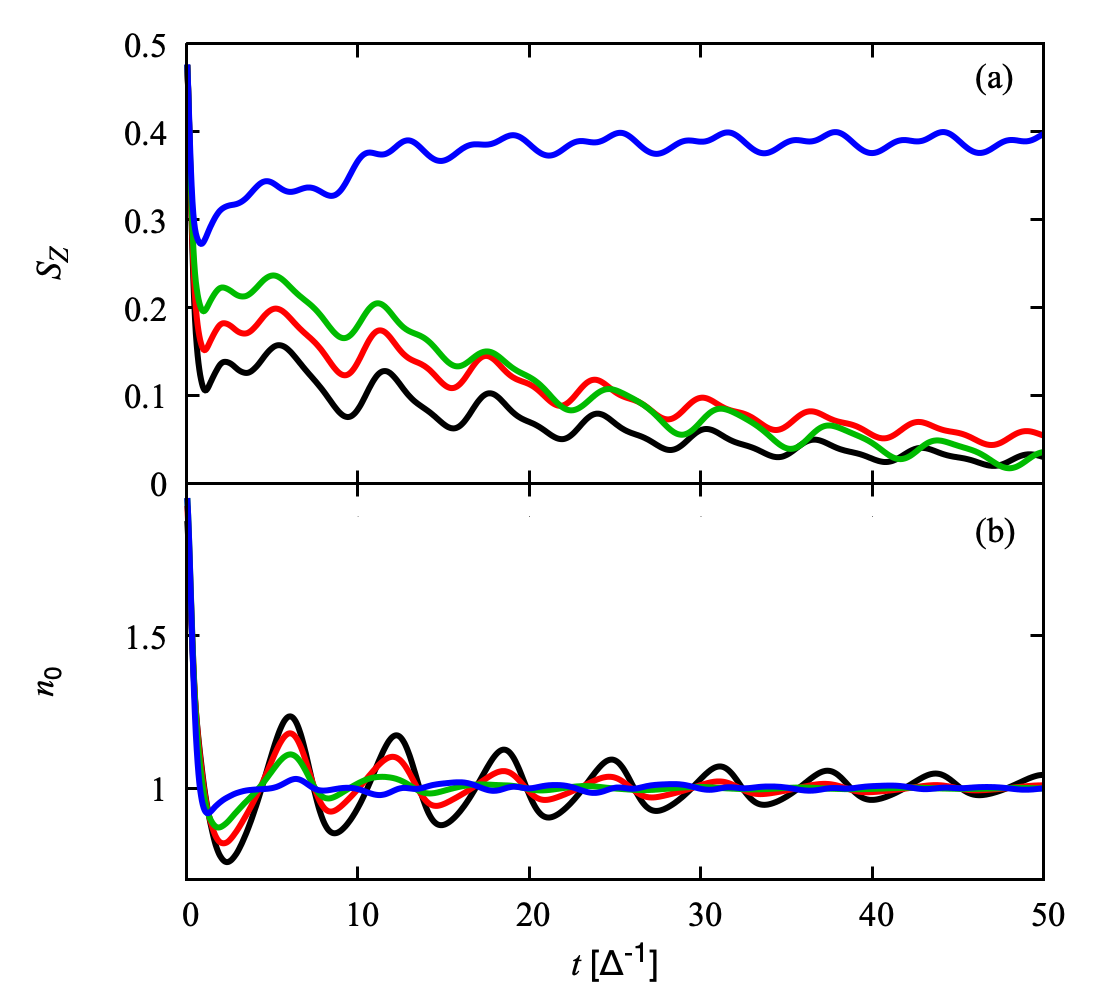}
    \caption{(a) Mean value of impurity spin polarization for the $(0,1)$ initial condition. (b) Impurity charge for the $(1,1)$ initial condition. We show results for the NEGFs with $V=\Delta/2$ and $U/\Gamma=0$ (black), $1$ (red), $2$ (green), and $4$ (blue). The remaining parameters are the same as in Fig. \ref{Fig:comparison_n01_V}: $\Gamma=\Delta=0.2t_0$, $\phi=0.6\pi$, and $\varepsilon_0=-U/2$.}
    \label{Fig:fig7}
\end{figure}

A weak electron-electron interaction does not modify the qualitative picture described above, Fig. \ref{Fig:fig7} (a). We note that Coulomb interactions tend to increase the relaxation rate with respect to the non-interacting situation in the $0$ phase, as shown by the green curve. The qualitative picture changes for strong Coulomb interactions, where  $S_Z$ converges to a non-zero long-time value, blue curve in Fig. \ref{Fig:fig7}. This change in the qualitative picture is related to the ground state transition between spin-singlet to spin-doublet \cite{Glazman_JETP1989,Rozhkov_PRL1999}. It indicates that the state with broken spin-rotation symmetry is a good stationary 
state of the system. As shown in App. \ref{Appendix:comparison_V}, this state is reached for an initially broken spin-rotation symmetry and $V/\Delta<1$, illustrating its robustness against perturbations. In contrast, for $V/\Delta\gtrsim2$ the $S_Z$ converges to a zero value, regardless of the initial condition.

In the case where the level is initially fully occupied, a finite bias voltage provides a relaxation mechanism for the impurity charge oscillations, Fig. \ref{Fig:fig7} (b). The charge converges to the expected steady-state in a time scale determined by $1/\gamma_T$ \eqref{gamma_T}.

\begin{figure}
    \centering
    \includegraphics[width=1\columnwidth]{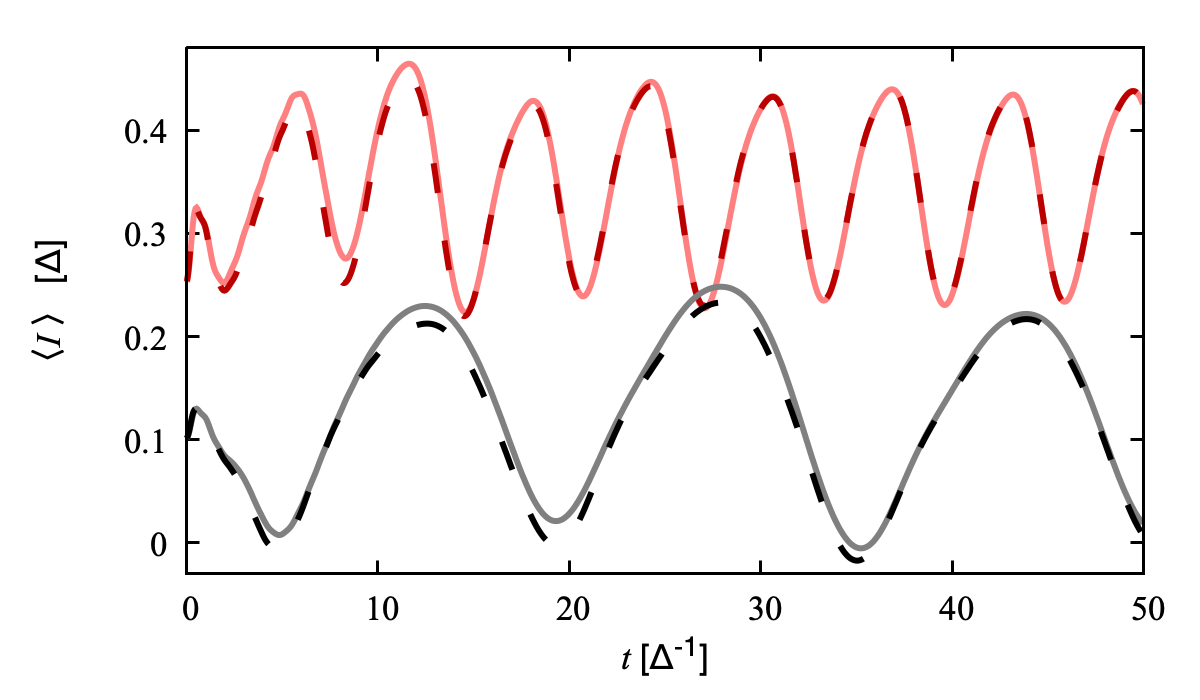}
    \caption{Symmetrized current through the junction for $V=0.2\Delta$ (black lines) and $V=0.5\Delta$ (red lines), for $U/\Delta=1$. We show results for $(0,1)$ (solid line) and $(1,1)$ (dashed lines) initial occupations. The remaining parameters are the same as in Fig. \ref{Fig:comparison_n01}: $\Gamma=\Delta=0.2t_0$, $\phi=0.6\pi$, and $\varepsilon_0=-U/2$.}
    \label{Fig:fig8}
\end{figure}

Finally, we analyze the non-equilibrium current through the impurity. We note that the symmetrized current, $\left\langle I\right\rangle=(I_L-I_R)/2$, shows a short time dependence on the initial conditions, Fig. \ref{Fig:fig8}. This is in contrast to the $\Gamma/\Delta\gg1$ limit, where the dependence on the initial condition is lost at very short times \cite{Souto_PRB17}. At sufficiently long times ($t\gg1/\Gamma_T$), the current converges to the same value, independently from the impurity initial condition. The current exhibits the characteristic time oscillations of the ac-Josephson effect, with a period given by $\pi/V$. At lower voltages, the dependence of the current on the initial conditions is visible at times much larger than any inverse energy in the model, illustrating the exponential dependence of the relaxation rate on the bias voltage, Fig. \ref{Fig:fig6} (a).

\section{Conclusions}
\label{Sec::Conclusions}
In this work, we have studied the non-equilibrium transient dynamics of an impurity tunnel-coupled to superconducting leads in the presence of local Coulomb interactions. We have developed two theoretical tools to study the time evolution of single-particle observables after a quench of the tunneling amplitudes. We have extended the exact time-dependent density matrix renormalization group (tDMRG) to the case where the leads feature superconducting correlations. We have furthermore developed an approximate method based on the second-order perturbation expansion in the interaction strength to study how the system approaches the steady-state. These two methods agree in the limit of weak to intermediate Coulomb interaction strengths. 

We have shown that the system gets trapped in an excited state for generic initial conditions, characterized by a lower supercurrent, similarly to the non-interacting case \cite{Souto_PRL16} for the non-interacting case. Local Coulomb interactions are not sufficient to make the system relax to the expected stationary state. We have also considered two additional relaxation mechanisms: (i)  coupling the impurity to a normal electrode, where excited quasiparticles can escape, and (ii)  a finite bias voltage, which allows quasiparticles to tunnel to the continuum of states where they can diffuse. Both mechanisms predict a spin-degenerate steady-state in the $0$ phase. In the $\pi$ phase, the system converges to a steady-state with finite spin-polarization for an initial condition with broken spin-rotation symmetry.

The methods developed in this work open the possibility of studying several interesting problems. The numerically unbiased tDMRG method is suitable to analyze the role of extended interactions in the leads and several coupled impurities under non-equilibrium conditions. This last situation is of current interest for the development 1-D topological devices \cite{Feldman_Nat2017}. On the other hand, the perturbative method is suitable for studying the long-time dynamics under generic non-equilibrium conditions. The method can be also extended to including topological superconducting leads \cite{Nilanjan_PRB2019,Souto_PRB2020} or additional impurities and leads, which can be either normal or superconducting.

\section{Acknowledgments}
RSS acknowledges funding from the European Research Council (ERC) under the European Union’s Horizon 2020 research and innovation programme under Grant Agreement No. 856526 and  from QuantERA project ``2D hybrid materials as a platform for topological quantum computing''. AEF is supported by the US Department of Energy (DOE), Office of Science, Basic Energy Sciences grant number DE-SC0019275. ALY and AMR acknowledge support from the Spanish MICINN through grants FIS2017-84860-R and through the “Mar\'ia de Maeztu” Programme for Units of Excellence in R\&D (Grant No. MDM-2014-0377) and by EU through grant no. 828948 (AndQC).

\bibliographystyle{apsrev4-2}
\bibliography{bibliography}

\clearpage
\appendix

\section{Current-phase relation}
\label{CPR}
\begin{figure}
    \centering
    \includegraphics[width=1\columnwidth]{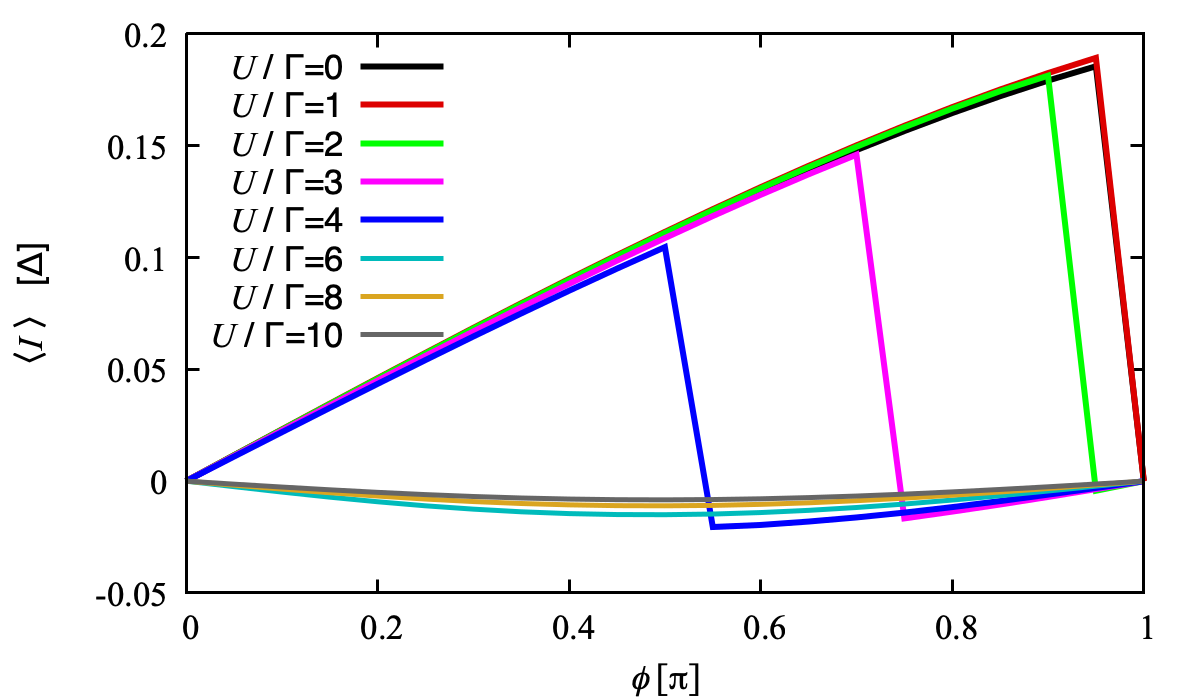}
    \caption{Current-phase relation for a phase-biased junction calculated using tDMRG. We use the same parameters as in Fig. \ref{Fig:comparison_n01}: $\Gamma=\Delta=0.2t_0$ and $\varepsilon_0=-U/2$.} 
    \label{Appendix:IvsPhi}
\end{figure}
In this section, we show the current-phase relation for some $U/\Gamma$ values using exact DMRG calculations, Fig. \ref{Appendix:IvsPhi}. In the non-interacting situation, the current is always positive. It exhibits a non-sinusoidal shape due to the high transmission chosen. In contrast, for $2<U/\Gamma<4$ we find a sudden transition of the current from a positive to a negative value, associated to the ground state change from singlet to doublet. Finally, for larger values of the interaction strength, the current is reversed with respect to the $U/\Gamma=0$ case, showing a sinusoidal dependence on $\phi$.

\section{Additional comparisons between tDMRG and NEGFs}
In this section, we show some results comparing the two develop methods. We include results for the evolution of the pair amplitude and the evolution of $S_Z$ with different bias voltages.

\subsection{Pair amplitude}
\label{Appendix:pair}
\begin{figure}
    \centering
    \includegraphics[width=1\columnwidth]{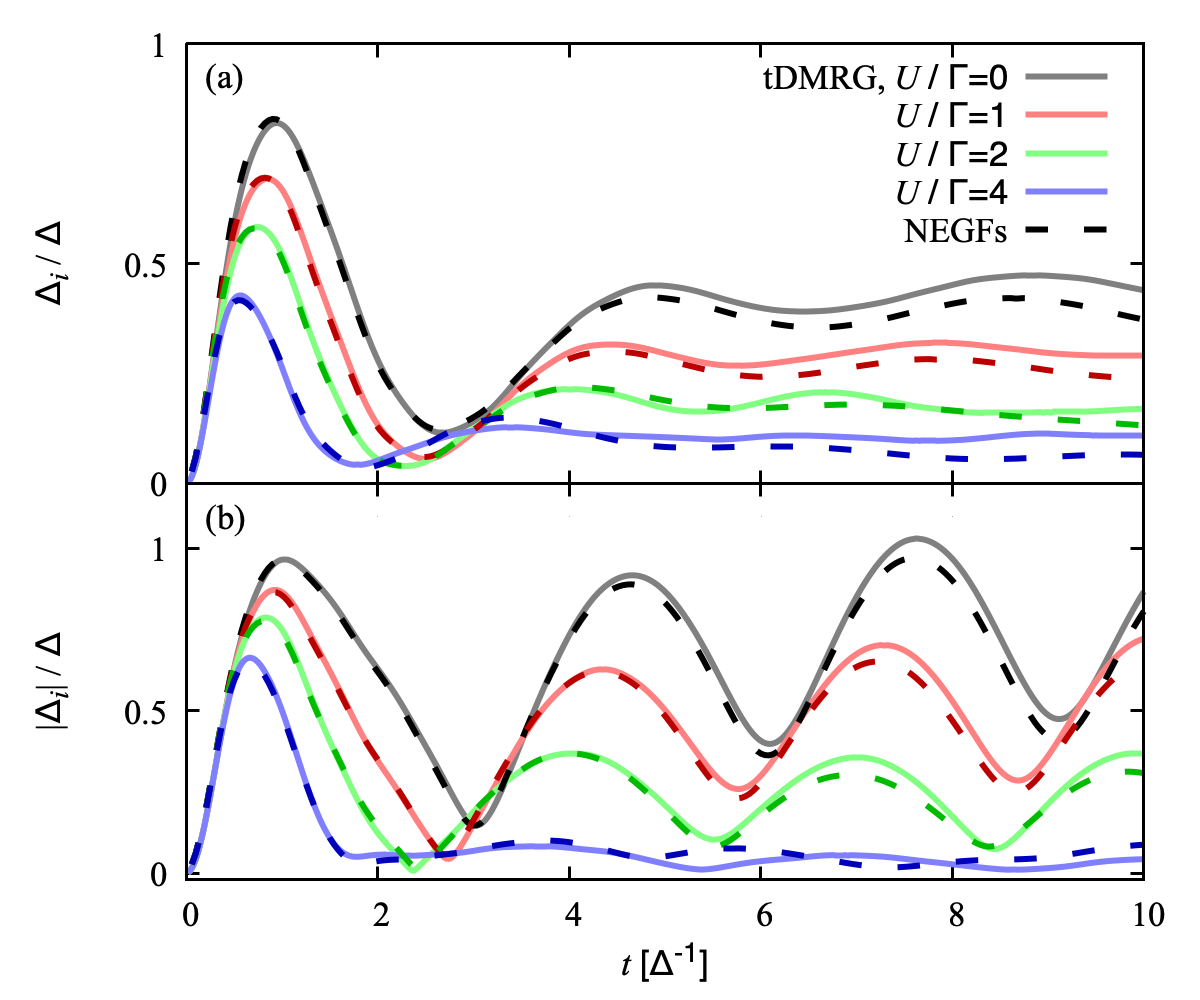}
    \caption{Pairing amplitude at the impurity for $V=0$ and $\phi=0$ and an initial occupation $(0,1)$ (a) and $(0,0)$ (b). We compare tDMRG (solid lines) with perturbative NEGFs (dashed lines). The remaining parameters are the same as in Fig. \ref{Fig:comparison_n01}: $\Gamma=\Delta=0.2t_0$, $\varepsilon_0=-U/2$, and $U/\Gamma=0$ (black), 1 (red), 2 (green), and 4 (blue).}
    \label{App:Delta}
\end{figure}

In Fig. \ref{App:Delta} we show the time evolution of the pair amplitude at the impurity after the connection for the two initial charge configurations: (0,1) and (0,0). In both cases, the exact tDMRG method and the approximated perturbative expansion using NEGFs agree remarkably. Coulomb interaction tends to damp the pair amplitude in the system, becoming nearly zero for $U/\Gamma\gg0$. This is due to the increased energy cost of having two electrons in the system due to the interaction. We note some differences between both methods for $U/\Gamma=0$, generated by the renormalization of the order parameter in tDMRG and absent in the NEGFs perturbative calculation.

The situation becomes richer for an initially occupied or empty impurity level, where the imaginary part of the order parameter oscillates in time around zero value. The real part converges to a stationary value, similarly to Fig. \ref{App:Delta} (a). The oscillation frequency is determined by the energy of the subgap state formed at the interface between the impurity and the superconductor \cite{Souto_PRL16,Souto_PRB17}. These oscillations are undamped for any value of $U/\Gamma$. However, their amplitude decrease when increasing Coulomb interaction as the energy difference between the ground state, with an electron in the impurity, and the excited states with a fully occupied or empty level increases. 

\begin{figure}
    \centering
    \includegraphics[width=1\columnwidth]{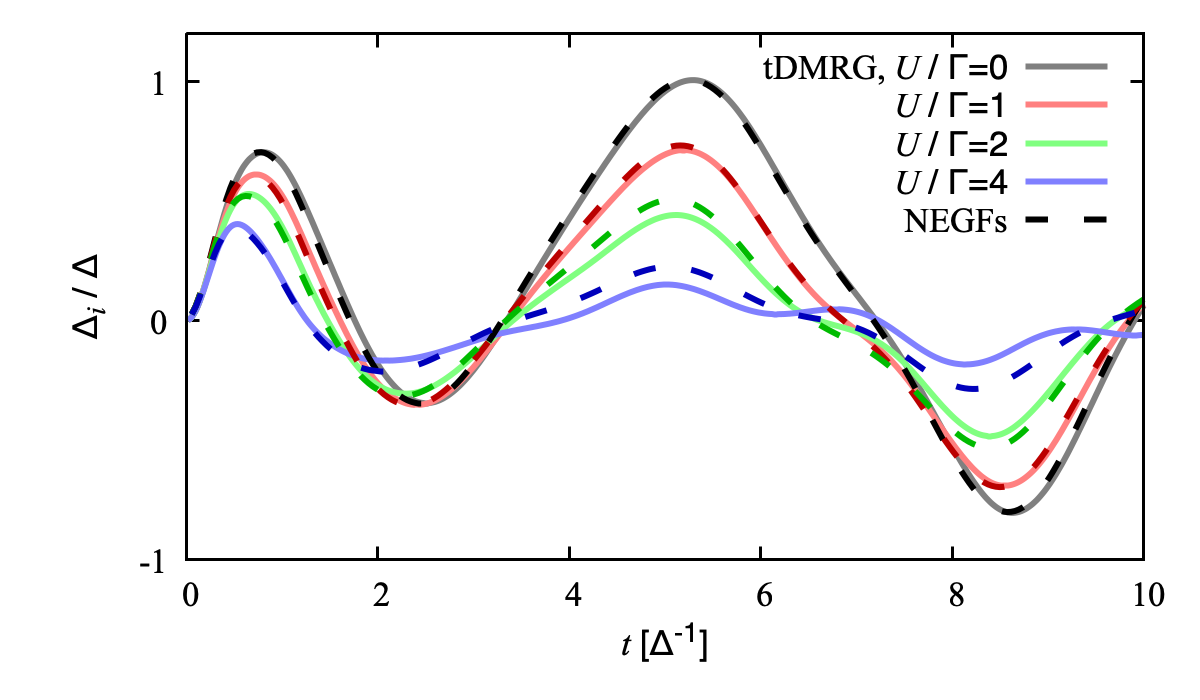}
    \caption{Pairing amplitude at the impurity for $V=\Delta$. The remaining parameters are the same as in Fig. \ref{Fig:comparison_n01}: $\Gamma=\Delta=0.2t_0$, $\varepsilon_0=-U/2$, and $U/\Gamma=0$ (black), 1 (red), 2 (green), and 4 (blue).} 
    \label{App:Delta_V1}
\end{figure}

In the bias voltage situation, the pair amplitude oscillates with a period given by the Josephson frequency, $V/\pi$, Fig. \ref{App:Delta_V1}. This is consistent with the fact that the phase difference in the leads increases linearly in time. We note that the shape of $\Delta_i$ is not sinusoidal due to non-adiabatic effects on the time evolution of the system. Similar to the phase-biased situation, $\Delta_i$ decreases with the interaction due to the enhanced energy cost of the doubly occupied (or vacant) state in the impurity.

\subsection{Finite bias voltage}
\label{Appendix:comparison_V}
\begin{figure}
    \centering
    \includegraphics[width=1\columnwidth]{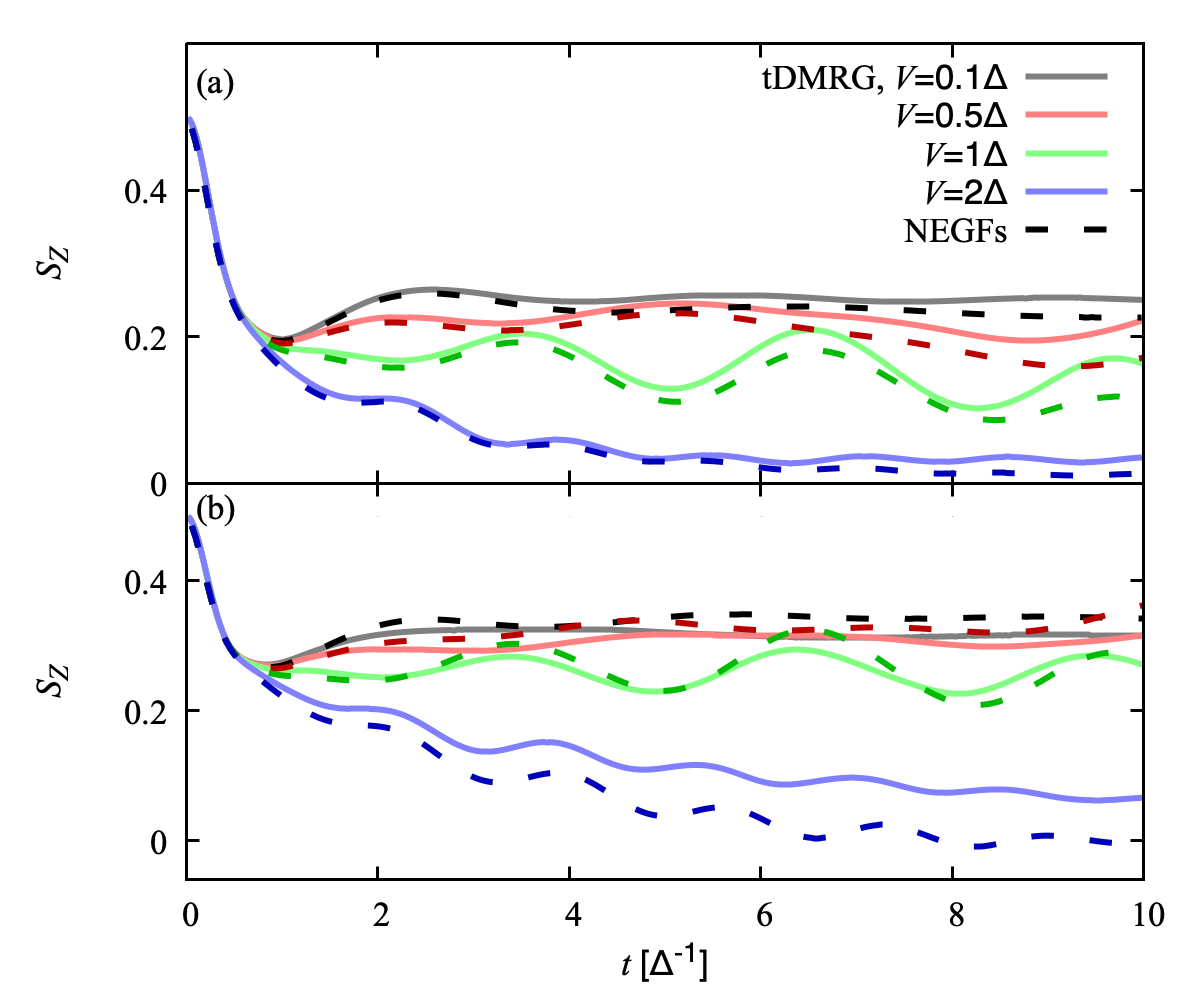}
    \caption{Magnetization evolution for $U/\Delta=2$ (a) and $4$ (b), and several voltage values. We compare tDMRG (solid lines) with the second-order approximation (dashed). The remaining parameters are the same as in Fig. \ref{Fig:comparison_n01}: $\Gamma=\Delta=0.2t_0$ and $\varepsilon_0=-U/2$.} 
    \label{App:Sz_V}
\end{figure}

In this section, we show additional results comparing tDMRG with the perturbative NEGF method for a voltage-biased junction. We show that both methods agree for a relatively wide range of parameters. They start to deviate for bias voltages larger than the gap and $U/\Gamma=4$, Fig. \ref{App:Sz_V} (b). However, both methods seem to converge to similar stationary values, approaching $S_Z=0$ for $V>\Delta$.

\end{document}